# Electron magnetic resonance in magnetic nanoparticles: dependence on the particle size and applicability of the modified giant spin model

N. Noginova[1*], B. Bates[1], V.A. Atsarkin[2]

[1]*Norfolk State University, Norfolk, Virginia, USA*
[2]*Kotel'nikov Institute of Radio Engineering and Electronics, Russian Academy of Science, Moscow, Russia*
*email: nnoginova@nsu.edu

**Abstract**
Electron magnetic resonance is experimentally studied in dilute solid suspensions of iron oxide nanoparticles as the function of the particle size, and discussed in the frames of the modified giant spin approach. Gradual evolution of features specific for small nanoparticles, including a narrow component at the main resonance field and a weak half-field line, is observed with the increase in the particle size, manifesting the transition from quantum to fully classical behavior. The shape, width, position of the resonance spectra, and the specific features are described quantitatively with a single set of fitting parameters for nanoparticles of 5-20 nm size. Limitations of the quantum model at high spin values are discussed.

## 1. Introduction

Magnetic nanoparticles (MNPs) attract a great interest due to various applications in a vast area from spintronics [1] to medicine [2]. This requires adequate quantitative description of the MNP features, and particularly the electron magnetic resonance (EMR) spectrum, which contains an important information on spin dynamics. Systems with small number of electron spins, such as individual spins of paramagnetic impurity or spin clusters with several spins coupled by strong exchange interactions, including molecular nanomagnets (MNMs), are commonly described with quantum approach. By increasing a number of coupled spins, one can expect a transition from purely quantum to classical behavior, with strongly different thermodynamics and a different shape of the electron magnetic resonance (EMR) spectrum. Magnetic nanoparticles containing hundreds or thousands of coupled spins are on the boundary between quantum and classical regimes. Both Zeeman and anisotropy energies of a single-domain MNP are comparable with the thermal energy $k_B T$, resulting in a significant role of thermal fluctuations in the magnetization dynamics of MNPs. Electron magnetic resonance in such superparamagnetic nanoparticles is often described using the classical approach based on Landau-Lifshits equations for ferromagnetic resonance (FMR) taking into account rapid thermal fluctuations of MNP magnetic moments in the framework of rotary diffusion [3-9]. However, there are features in the EMR spectrum of small MNPs [10-17], which are not readily described by the classical approach, including a narrow component at the main resonance field, $B_0$, and transitions at $B_0/n$, $n$ = 2, 3, 4. Such features are typical for electron paramagnetic resonance of single spins but not for classical FMR.

Recently the "giant spin" approach was employed for description of the EMR properties of small MNPs [10]. In this approach, a nanoparticle is considered as a giant spin, $S$, originated from exchange coupled electron spins (equivalent to the magnetic moment μ of the nanoparticle). The EMR spectrum is the sum of the contributions from resonance transitions between separate energy levels $E_m$, corresponding to projections $m$ of the spin $S$ onto direction of the quantization axis (in the high-field approximation, the direction of the external magnetic field $B$). The thermal scatter of µ directions is reflected in the Boltzmann distribution of populations over the energy levels $E_m$. Unlike classical description, this approach can satisfactory explain the presence of the narrow component in the main resonance lineshape as well as its temperature dependence. Besides, this quantum-based approach describes additional features at $B_0/n$ in EMR spectra of nanoparticles in similarity with forbidden resonance transitions typical for paramagnetic spin systems [11-13]. In Refs. [14-17], a thorough comparative analysis of the MNP and MNM spectra was performed in the framework of the giant spin model, thus bridging the gap between the two classes of nano-objects.

At the same time, the giant spin model presented in [10] is a significant simplification. It neglects upper spin multiplets with $S$-1, $S$-2, etc. (this point was specially discussed in [14, 17]). A number of phenomenological parameters has to be introduced in order to provide quantitative fitting of the experimental EMR spectra. In addition, this consideration does not explain a steep decrease of "multiple quantum transition" magnitudes with cooling [11, 13] and does not provide criteria for a possible transition of MNPs to fully classical behavior with the decrease of temperature and increase in the MNP size.

In order to get more information on the transition from quantum to classical type of behavior in spin systems, the goal of this work is to experimentally study the evolution of the EMR spectra as the function of the particle size. Particular attention is paid to the evolution of "quantum" features with the increase in the particle size, in particular, a narrow central component and forbidden quantum transition signals observed at the half of the main resonance field. We also make an attempt to fit the experimental data in frames of the giant spin approach.

## 2. Experiment

In our experiments we used iron oxide nanoparticles (NP) with sizes ranging from 5 nm to 40 nm. Several series of $Fe_3O_4$ nanoparticles with oleic acid as a surfactant were obtained from the same source, NN-Labs [18]. Main experiments and modeling were performed with relatively small particles, including: NP 5 with the average diameter, $d$, of 5 nm ± 1 nm; NP 9: $d$ = 9.5 nm ± 1 nm; NP 10: $d$ = 10.5 nm ± 1 nm; NP 20: $d$ = 20 nm ± 3 nm. Series of $Fe_3O_4$ with the average size of 30 nm and 40 nm were used for comparison in order to see the further trend. According to the characterization data provided by the manufacturer, the magnetization of nanoparticles was 45 emu/g for NP 5, 30 emu/g for NP 9 and NP 10, and 20 emu/g for larger particles (measured at 4.5 T and room temperature). In addition to $Fe_3O_4$, in our study we included a different sample (NP 15), which appeared to show a similar behavior, core-shell ($\gamma$-$Fe_2O_3$:Au) nanoparticles with oleylamine on the surface, with the average size of ~ 15 nm ± 2nm (fabricated at Cornell University).

The experimental samples were solid suspensions of nanoparticles in polymer, prepared by mixing toluene solutions of polystyrene and ferrofluid, and subsequent drying off the solvent. The samples were well diluted to ensure that in a dry system, the particles were separated from each other in order to exclude an additional broadening due to dipolar interactions. This was tested by a further decrease in the relative

concentration of NP and observing possible changes in the linewidths of the main EMR and the narrow component

The EMR measurements were performed using a standard Bruker EPR spectrometer at ambient temperature. The microwave frequency was 10 GHz; the modulation frequency was 100 kHz. The EMR signal measured by a standard EPR spectrometer presents the derivative of the microwave absorption over the magnetic field. The EMR spectra of all series are shown together in Fig. 1. One can see a typical gradual evolution of the spectra with an increase in size: the main signal becomes broader and shifts toward lower fields. In order to compare the spectra of different series in further analysis, the signal for each sample was normalized to the corresponding double integral. The central part of the spectra is shown in Fig. 2 (a). The narrow component with g ~ 2 is seen as a dominant feature in the EMR of NP 5, and can also be distinguished in spectra of 9 -10 nm series. In order to check a possible presence of the narrow component in the EMR of larger sizes, the derivative *dI/dB* of the EMR signal (which corresponds to the double derivative of the magnetically dependent microwave absorption over the magnetic field) is plotted in Figs. 2 (b) and (c). The narrow component is seen as a dip in *dI/dB* at $B_0$ ~ 3440 G with the depth decreasing with the growth of *d*. A shallow dip is observed in the NP 15. No presence of a dip is distinguished for *d*= 20 nm and larger.

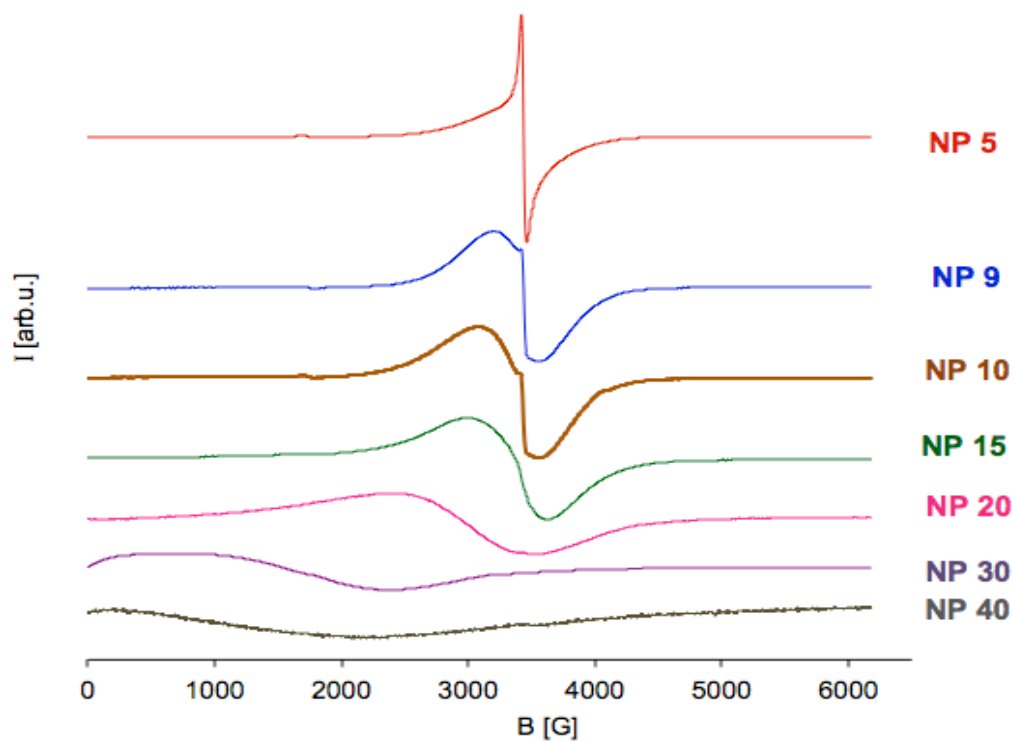

**Figure 1**. Shapes of the EMR spectra in iron oxide nanoparticles of various sizes. The samples were indicated.

The signal at the half-field (see a small feature at ~ 1750 G in the NP 5 spectrum, Fig. 1), is considered as the "forbidden" transition in frames of the quantal approach [11,12]; it corresponds to the doubled change in the spin projection, $\Delta m = \pm 2$, so we denote it as 2Q (double quantum). This line can be extracted from the EMR spectrum by fitting the low-field wing of the main signal with a smooth broad curve (a combination of Gaussian and Lorentzian lines.) The 2Q signal in NP 5, NP 9 and NP 15 samples is plotted in Fig. 2(d). For comparison, the main resonance for the same sample is shown in Fig. 2(e).

Similarly to the main resonance, the half-field signal consists of the broad and narrow components, however, here the narrow component is more pronounced. Note that the abscissa scale of Fig. 2(e) is twice larger than that of Fig. 2(d). Thus, seemingly of the same width, the broad component in the 2Q resonance is approximately twice narrow than the broad component of the main resonance while the widths of the narrow components in both resonances are of the same order.

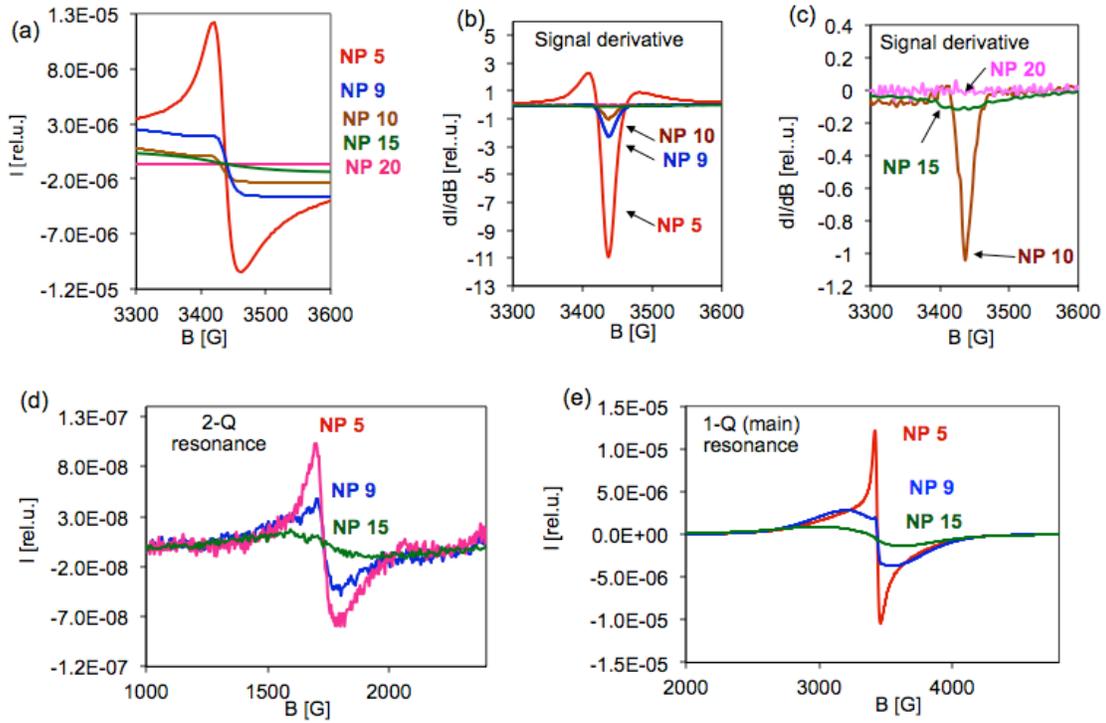

**Figure 2**. (a) The central part of the normalized EMR spectra $I(B)$; (b) and (c) The derivative $dI/dB$ indicating the presence of the narrow component; (d) Half-field and (e) main resonance signals. The samples are indicated.

Let us estimate the contributions of the narrow component and the half-field signal to the total signal. The relative weight $W_N$ of the narrow line can be found by two methods. First, fitting the broad signal of the main resonance by a combination of Gaussian and Lorentzian lines near the central field, we extract the narrow signal, and numerically double integrate the result. In the other method, the weight of the narrow signal is estimated as a product $A*(\delta_{narrow})^3$ where $A$ is the depth of the dip in the derivative of the EMR signal (Fig. 2(b,c)) and $\delta_{narrow}$ is the effective width (~40 G) of the narrow signal. The relative weight of the half-field signal $W_{2Q}$ is estimated through the double integration of the extracted signal (Fig. 2 (d)).

As one can see, $W_N$ quickly decreases with the growth of $d$. The contribution $W_{2Q}$ changes weekly and non-monotonously with the increase of $d$ from 5 nm to 15 nm. No half-field signal can be resolved for larger MNPs.

## 3. Fitting and Discussion

In the numerical simulations, we use the quantization approach described in detail in [10-12]. Following [10], the magnetic moment of a single-domain nanoparticle is considered as a "giant spin", $S$. In the high-field approximation, when the external magnetic field $B$ exceeds strongly both the anisotropy and demagnetization fields, the spectrum can be calculated as a sum of the contributions from transitions between energy levels $E_m$ corresponding to the projection, $m$, of the total spin onto $B$ direction in a way similar to calculations of electron paramagnetic resonance spectra for paramagnetic impurity ions. We take into account the probabilities, $W_m$, of the allowed ($m, m+1$) transitions as [19]

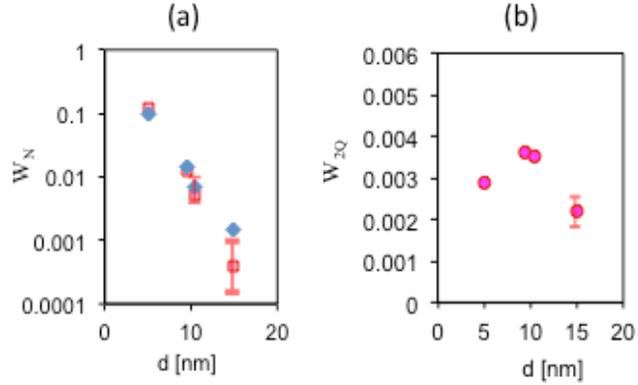

**Figure 3**. Evolution of the "quantum features" with the particle size: (a) the relative weight of the narrow signal estimated through the signal extraction (squares) and the derivative (diamonds), and (b) relative weight of the half-field resonance.

$$W_m = C\, g_m(B\text{-}B_m)[S(S+1) - m(m+1)], \quad (1)$$

and the Boltzmann distribution of populations at the magnetic sublevels,

$$\rho_m = Z_\rho^{-1} \exp(\frac{-E_m}{k_B T}), \quad (2)$$

where $C$ is the proportionality factor, $g_m(B\text{-}B_m)$ and $B_m$ are the form-factor and resonance field of the resonance line at the transition involved, and $Z_\rho$ is the partition function. Assuming the high-field condition, we approximate the energy level $E_m$ in Eq.(2) by its Zeeman term, $-\gamma\hbar m B$, where $\gamma$ is the electron gyromagnetic ratio.

Following [10], we account for a shift in the resonance field to lower fields commonly observed with a decrease in temperature or an increase in the particle size. The shift may originate from surface spins or the second order of the anisotropy effects. In both cases, the magnitude of the shift depends on $m$ [10]. In our simulations, it is described phenomenologically as

$$B_m - B_0 = -k_s|m|, \quad (3)$$

where $k_s$ is the fitting parameter.

As discussed in [10], the effect of the uniaxial anisotropy in nanoparticles can be compared with effect of the crystal field in the paramagnetic case, which shifts the resonance field of a given transition *(m, m+1)* by a factor proportional to *m*. To calculate the overall EMR spectrum of the MNP suspension, the integration should be performed over the angles $\theta$ between *B* and randomly distributed directions of the individual anisotropy axes. This results in asymmetric "powder" spectrum, which generally contradicts to the experiment, where the broad component of the EMR line becomes more symmetric with increase in the particle size or decrease in temperature. Note that the assumption of uniaxial anisotropy of nanoparticles used in [10] was an idealization, which could not fully describe particles of a random shape. Let us now take into account a broadening of the transitions as well, which can be associated with various dynamic effects and static spread in anisotropy (surface- or shape-related), and assume that this broadening grows with an increase in *m* as well. In our fitting we use a simple phenomenological expression for a contribution from a particular transition *(m, m+1)*

$$\delta_m = \delta_0 + \delta_{FMR} (1 - \exp(-|m|/b)), \quad (4)$$

where $\delta_0$ is the width of the central transition (-½, ½), $\delta_{FMR}$ is the linewidth of the classical FMR and *b* is the fitting parameter. The origin of the broadening may include fast fluctuations leading to an acceleration of the spin relaxation, in analogy to the acceleration of high-spin NMR transitions due to fluctuations of the electric field gradient [20] or slow fluctuations, affecting the transition frequency. Additional nonhomogeneous broadening results from chaotic distribution of the MNP anisotropy axes.

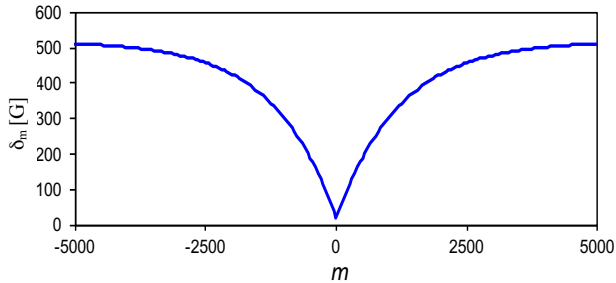

**Figure 4.** Width of *(m, m+1)* transition according to the model, Eq. (4).

According to Eq. (4) (Fig.4), the linewidth grows at small $|m|$ (as is expected in the paramagnetic case due to homogeneous and inhomogeneous broadening associated with the crystal field anisotropy) and saturates to $\sim \delta_{FMR}$ at large $|m|$. Note, that this saturation does not follow from our initial model [10], and was accepted in order to get the best fitting.

Particles used in our experiments differ in quality and have different magnetizations. However, here we make an attempt to fit all the samples using the same fitting parameters except of *S* which is estimated from the magnetization data.

We take into account that NP 15 nanoparticles had a ~ 1 nm gold layer on the surface, and model these particles as 13 nm sized. We find that the best fit can be obtained supposing the saturation magnetization of 150 - 200 emu/cm$^3$, which is significantly lower than that of the bulk material (>400 emu/cm$^3$), however, is in agreement with the characterization data for the series under study. Based on the magnetization data and best fitting, the magnetic moments of nanoparticles are estimated to be 2260, 7620, 10450, and 23000 Bohr magneton for NP 5, NP 9, NP 10, and NP 15 respectively, which corresponded to spherical particles with the magnetization of 185 emu/cm$^3$ and the diameter of ~ 6 nm, 9 nm, 10 nm and 13 nm. In the fitting shown in Fig. 5(a), the form-factor of the partial transition is presented with the sum of the Lorentzian and Gaussian curves (with the relative weights 0.6:0.4). Parameters used in this fitting are, $k_s$ = 0.08 G, $\delta_o$ = 20 G, $\delta_{FMR}$ = 500 G and $b$ = 1200. One can see that the observed spectra are in general agreement with the calculated ones, except for some asymmetry, which originates from a random distribution of the anisotropy axes and is not taken in account in frames of our simplified model.

According to [11], there are two main sources of 2Q transitions: dipolar interactions between particles and an anisotropy of a particle. In a well-diluted system, only the anisotropy plays a role. The probability of the transitions, $W_{2Q}$ can be calculated as

$$W_{2Q} = (B_a^{eff}/B_0^{2Q})^2 \, S^2 \, [(1-(m/S)^2)]^2, \quad (5)$$

where $B_0^{2Q} = B_0/2$ is the resonance field of the 2Q transitions, and the effective anisotropy field, $B_a^{eff}$, is the fitting parameter.

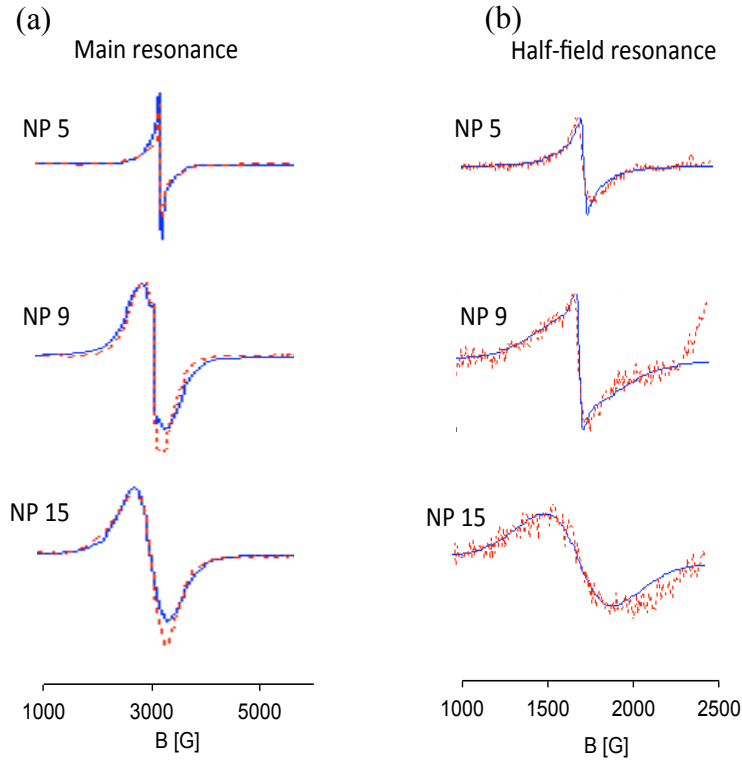

**Figure 5**. Fitting the main resonance (a) and half-field signal (b) with the giant spin approach, experiment (red) and calculations (blue).

In the simulations of the 2Q signal, Fig. 5(b), we follow the procedure similar to the calculations of the main signal. One can assume that the linewidth is expected to be twice smaller in the half of the resonance field. We find that the experimental curves can be satisfactory fitted with the parameters $\delta_o = 20$ G and $\delta_{FMR} = 300$ G and $b = 600$, which was close to the predictions. However, in opposite to the main resonance, no shift of the 2Q signal is observed with the growth in the particle size, so that $k_s$ was assumed to be 0 in the simulations of 2Q signal. The effective anisotropy field used in the fitting, $B_a^{eff}$ =100 G, which seems to be quite reasonable.

In Fig. 6, the relative weights of the narrow signal and the half-field transition are compared with the calculated values. As one can see, the simulations adequately describe the size dependence of these two features and predict their disappearance with further increase in $S$.

As seen from Figs. 5, 6, the calculations based on the simplified giant spin approach well agree with the experimental data for both the main resonance line and the half-field signal. It should be noted that the data obtained in the whole range of MNP sizes, from 5 to 15 nm, were fitted using a single set of fitting parameters (except, of course, the $S$ values determined from the MNP magnetic moments). Success of the model seems to be even surprising, keeping in mind the accepted approximations, and especially the replacement of the angular integration over all possible angles of the anisotropy axis orientation by a sum of isotropic Gaussian-Lorentzian partial lines of quantum transitions. Inaccurate accounting for the axial magnetic anisotropy can be a source of some discrepancy between experiment and calculations in the shape of the main resonance seen in Fig.5(a), (NP 9 - NP 15).

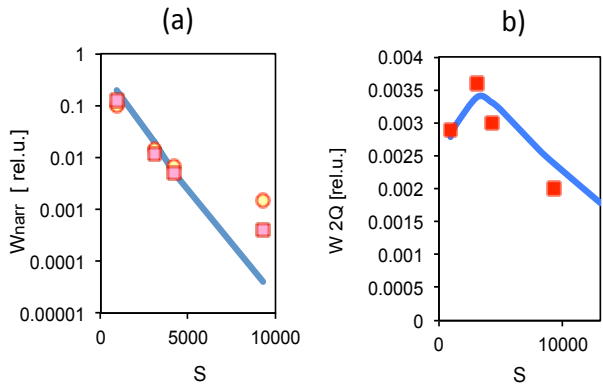

**Figure 6**. Relative weights of the narrow component (a), and half-field signal (b) as the functions of $S$. Points are the experimental data, solid traces are calculations.

Another unclear assumption made in our fitting is the restriction of the line broadening at $|m|>b$, see Eq.(4). We find that without this limitation, a satisfactory fitting appears to be impossible. Probably, it relates with the inherent defect of the present giant spin model accounting only the lowest spin multiplet $S$ and neglecting the contributions from the excited multiplets $S$-1, $S$-2,…, which can be well populated at room temperature. This problem was discussed in Ref. [17] where some effect of the upper spin multiplets on the MNP spectra was observed experimentally. In any case, the assumed saturation of the transition broadening, Fig. 4, is justified by the fact that the classical FMR line-width in macroscopic objects does not depend on $S$. Thus, the saturation of the linewidth magnitude at high values of $|m|$ may be considered as a limitation of the simple giant spin model and an indication of the transition to the classic approach with the corresponding averaging [3, 4].

An important finding of this work is another evidence for the quantum nature of the central narrow EMR component which appears quite naturally originating from the transitions with low $|m|$. The relative weight of this component quickly decreases with the increase of the system size. This fact is readily explained by our approach with the decrease in the population of central levels (corresponding to low |*m*|) with growing *S*. It should be noted, however, that there exists other interpretations of the narrow feature. One of them is suggested and widely used by the Berger – Kliava group (see, for example, Refs.[5-7]). Their explanation accounts for a distribution of the MNP sizes in a given sample. As a result, small enough particles with strongly averaged anisotropy form the narrow feature under discussion. Another approach relates the narrow central line to a surface layer, which has the EMR spectrum differing from that of the inner (volume) part of the particle [21]. Both interpretations are able to explain, at least qualitatively, the temperature and size dependences of the narrow component and may be considered as possible alternatives to the giant spin approach. Note, however, that the latter has no need in additional objects, such as smaller particles or a surface layer.

In conclusion, the evolution of EMR spectra of magnetic nanoparticles with the particle size has been studied in well-diluted solid suspensions of iron oxide nanoparticles. A sharp decrease in the relative magnitude of the narrow component was observed with an increase in the particle size. The half-field signal demonstrated weak size dependence at small nanoparticles and disappeared with further increase in the particle size. The experimental curves and dependences can be satisfactory fitted with the modified model based on the giant spin approach.

**Acknowledgments**
The work was partially supported by PREM grant no. DMR 1205457 and the Army Research Office (ARO) Grant W911NF-14-1-0639 V.A. was supported by the Grant from Russian Foundation for Basic Research (RFBR) 14-02-00165. Authors thank E. Giannelis for providing the samples.